\documentclass[12pt,a4paper,notitlepage]{article}

\usepackage[utf8]{inputenc} 
\usepackage[T1]{fontenc}    
\usepackage{hyperref}       
\usepackage{url}            
\usepackage{booktabs}       
\usepackage{amsfonts}       
\usepackage{nicefrac}       
\usepackage{graphicx}
\usepackage{amsmath,amssymb}
\usepackage{mathrsfs}
\usepackage{MnSymbol}
\usepackage{bbold}
\usepackage[margin=0.8in]{geometry}
\usepackage{hyperref}
\hypersetup{
 colorlinks=true,
 citecolor=blue,
 linkcolor=blue,
 urlcolor=blue,
 pdfpagemode=UseNone,
 pdfstartview=FitH}

\providecommand{\bra}[1]{\langle #1 \rvert}
\providecommand{\ket}[1]{\lvert #1 \rangle}
\providecommand{\iprod}[2]{\langle #1\vert#2 \rangle}

\DeclareMathOperator{\real}{Re}
\DeclareMathOperator{\spn}{span}
\DeclareMathOperator{\trace}{tr}

\begin{document}

\title{Entanglement entropy distinguishes \\ 
PT-symmetry and topological phases \\
in a class of non-unitary quantum walks}

\author{
 Gene M.~M.~Itable$^*$ and Francis N.~C.~Paraan$^{**}$  \\
  National Institute of Physics\\
  University of the Philippines Diliman\\
  1101 Quezon City, Philippines \\
  $^*$\texttt{gitable@nip.upd.edu.ph}, 
$^{**}$\texttt{fparaan@nip.upd.edu.ph} \\
  }

\maketitle

\abstract{\noindent
We calculate the hybrid entanglement entropy between coin and walker degrees of freedom in a non-unitary quantum walk. The model possesses a joint parity and time-reversal symmetry or PT-symmetry and supports topological phases when this symmetry is unbroken by its eigenstates. An asymptotic analysis at long times reveals that the quantum walk can indefinitely sustain hybrid entanglement in the unbroken symmetry phase even when gain and loss mechanisms are present. However, when the gain-loss strength is too large, the PT-symmetry of the model is spontaneously broken and entanglement vanishes. The entanglement entropy is therefore an effective and robust parameter for constructing PT-symmetry and topological phase diagrams in this non-unitary dynamical system.}


\section{Introduction}\label{sec:intro}

In a discrete-time quantum walk \cite{nayak2000,ambainis2003}, the hopping amplitudes between base kets of a ``walker'' state space $\mathscr{H}_\text{w} = \spn ( \{ \ket{n} \} ), n \in \mathbb{Z}$  are conditioned on an internal ``coin'' state in $\mathscr{H}_\text{c} = \spn  ( \ket{L}, \ket{R} )$. The dynamics of the quantum state $\ket{\psi} \in \mathscr{H}_\text{w}\otimes \mathscr{H}_\text{c}$, given an initial state $\ket{\psi(0)}$, is given by the evolution equation
\begin{equation}
    \ket{\psi(t)} = \mathcal{U}^t \ket{\psi(0)},
    \label{eq:evolution}
\end{equation}
where the single step operator $\mathcal{U} = \mathcal{SR}$ typically consists of a shift operation
\begin{equation}\label{eq:shift}
    \mathcal{S} = \sum_{n}  \ket{n+1}\bra{n} \otimes \ket{R}\bra{R} + \ket{n-1}\bra{n} \otimes \ket{L}\bra{L}, 
\end{equation}
preceded by a generic operation $\mathcal{R}$ on the coin space, and $t$ is a non-negative integer denoting the number of steps in the walk. Sometimes, the step operator $\mathcal{U}$ consists of several shift and coin space operations, as in the multi-step quantum walk studied here.

A notable feature of quantum walks is the rapid spread of an initially localized walker, which can be used to engineer efficient quantum search and targeting algorithms \cite{ambainis2003}. These walks can also serve as frameworks for other quantum information tasks, such as universal quantum computation \cite{childs2009universal,lovett2010universal} and quantum-state transfer \cite{kurzy2011discrete}. Experimentally, quantum walks are often realized in photonic \cite{schreiber2010photons} and ultracold atomic setups \cite{karski2009quantum,preiss2015}, where a great degree of control, reproducibility, and tunability can be routinely achieved \cite{manouchehri2014}. Thus, they have been utilized as robust simulators for difficult-to-engineer Hamiltonians, such as those involved in the study of localization and topological phases in disordered systems and non-Hermitian physics \cite{denicola2014quantum,regensburger2012parity,xiao2017,zhan2017,ozdemir2019,hatano2021,lin2022}.  

As a quantum walk progresses and the wavefunction spreads over space, entanglement can be generated between the walker and coin spaces \cite{carneiro2005,abal2006,ide2011}. This hybrid entanglement involving distinct degrees of freedom is valuable because it can be used to store additional quantum information in internal state spaces \cite{neves2009hybrid,li2018hyper,flamini2018photonic}, and several strategies aimed at improving entanglement yield in unitary walks exist. For instance, optimizing coin sequences in multi-step walks \cite{gratsea2020,gratsea2020b} and applying dynamical noise on coin operations \cite{vieira2013} can maximize the coin-walker entanglement entropy in the long-time limit.

Since entanglement is fragile and quantum devices can not be fully isolated from their surroundings, it is important to know how much entanglement a system can have when decoherence is present. One way of introducing decoherence into a quantum walk is by generalizing the evolution operator to a non-unitary one, which is equivalent to having a non-Hermitian effective Hamiltonian. For instance, in quantum walks with randomly occurring projective measurements, decoherence can lead to the decay and vanishing of quantum correlations and entanglement between the coin and walker \cite{maloyer2007}. However, the current understanding of the relationship between entanglement and decoherence in non-Hermitian systems is quite model-dependent \cite{dey2019,fring2019,chakraborty2019delayed}, and there is no immediate indication whether entanglement can be maintained or not in a given system with decoherence. Indeed, we find here that hybrid entanglement can persist in a quantum walk even when the evolution is not unitary.  

In this work we study this relationship between entanglement and non-unitarity in a class of quantum walks in a parameter space that is divided into two regions: one that features eigenstates that possess a joint parity and time-reversal symmetry or PT-symmetry \cite{bender1998a,bender1999}, and another where this symmetry is spontaneously broken. In particular, the model \cite{mochizuki2016} is a two-step quantum walk that accounts for photonic gain, loss, and general coin rotations in optical setups \cite{regensburger2012parity,xiao2017,zhan2017} that have been used to investigate the deep connections between PT-symmetry and topological phases of matter (Section \ref{sec:model}).  Motivated by the fact that entanglement has been successfully used to probe various quantum phase transitions \cite{lambert2005,dechiara2012,wang2020}, we do the same by demonstrating that the long-time entanglement entropy can effectively identify the PT-symmetry phases of the model (Section \ref{sec:entanglement}). Moreover, since level crossings in this type of PT-symmetry breaking are associated with topological phase transitions \cite{mochizuki2020,wang2022}, we also show that the characteristics of the entanglement entropy are distinct in the different topological phases of the model (Section~\ref{sec:topo}).   

Our most significant and compelling findings are summarized in phase diagrams (Figure~\ref{fig:phase_diagram}) that show how entanglement can distinguish the different PT-symmetry phases of the model, and graphs (Figure~\ref{fig:gap}) that shows how an entanglement gap forms between topological phases of the model as the gain-loss strength and degree of non-unitarity is increased.

\section{Non-unitary quantum walk model}\label{sec:model}

We consider a spatially homogeneous two-step quantum walk so that a Fourier basis $\{ \ket{k} \}$ labeled by wavevectors $k$ in a Brillouin zone (BZ) such that
\begin{equation}
   \ket{\psi(t)} \equiv \frac{1}{\sqrt{\pi}}\int_{\text{BZ}} \ket{k}\otimes \ket{\tilde{\psi}_k(t)} \,\text{d}k,\quad \text{BZ}\in[-\pi/2,\pi/2),
\end{equation}
and the step operator $\mathcal{U}$ is partially diagonalized:
\begin{equation}
    \mathcal{U} = \frac{1}{\pi}\int_{\text{BZ}} \vert k \rangle \langle k \vert \otimes \tilde{U}(k)\, \text{d}k.
\end{equation}
The evolution operator $\tilde{U}(k)$, which acts only on the qubit coin space, has the specific form
\begin{equation}
    \tilde{U}(k) = \tilde{S}(k)\tilde{G}(\gamma)\tilde{\Phi}(\phi)\tilde{C}(\theta_2)\tilde{S}(k)\tilde{G}(-\gamma)\tilde{\Phi}(\phi)\tilde{C}(\theta_1),
\end{equation}
where $\tilde{S}(k) = e^{ik\sigma_3}$ is the Fourier transform of the shift operator $\mathcal{S}$ (Eq.~\ref{eq:shift}), $\tilde{G}(\gamma) = e^{\gamma\sigma_3}$ is a gain-loss operator, $\tilde{\Phi}(\phi) = e^{i\phi\sigma_3}$ is a phase-shift operator, $\tilde{C}(\theta_j) = e^{i\theta_j\sigma_1}$ is a coin rotation operator, and the $\sigma_j$'s are Pauli operators. The model parameters $\theta_j, \phi,$ and $\gamma$ are real so that the evolution is unitary when the gain-loss parameter is {$e^\gamma = \pm 1$} and non-unitary otherwise. The phase shift angle $\phi$ translates the origin of wavevector space and hereafter we make the replacements $k +\phi \to k$ and $\tilde{\Phi}(\phi) \to \mathbb{1}_k$ for simplicity. In photonic experimental realizations of this and similar quantum walks, the shift operators are typically implemented by beam displacers, the gain-loss operators by partially polarizing beam splitters, and the phase-shift and generic coin rotation operators by wave plates \cite{regensburger2012parity,xiao2017,mochizuki2020}. 

\begin{figure}[t]
\centering
\includegraphics[width=0.96\linewidth]{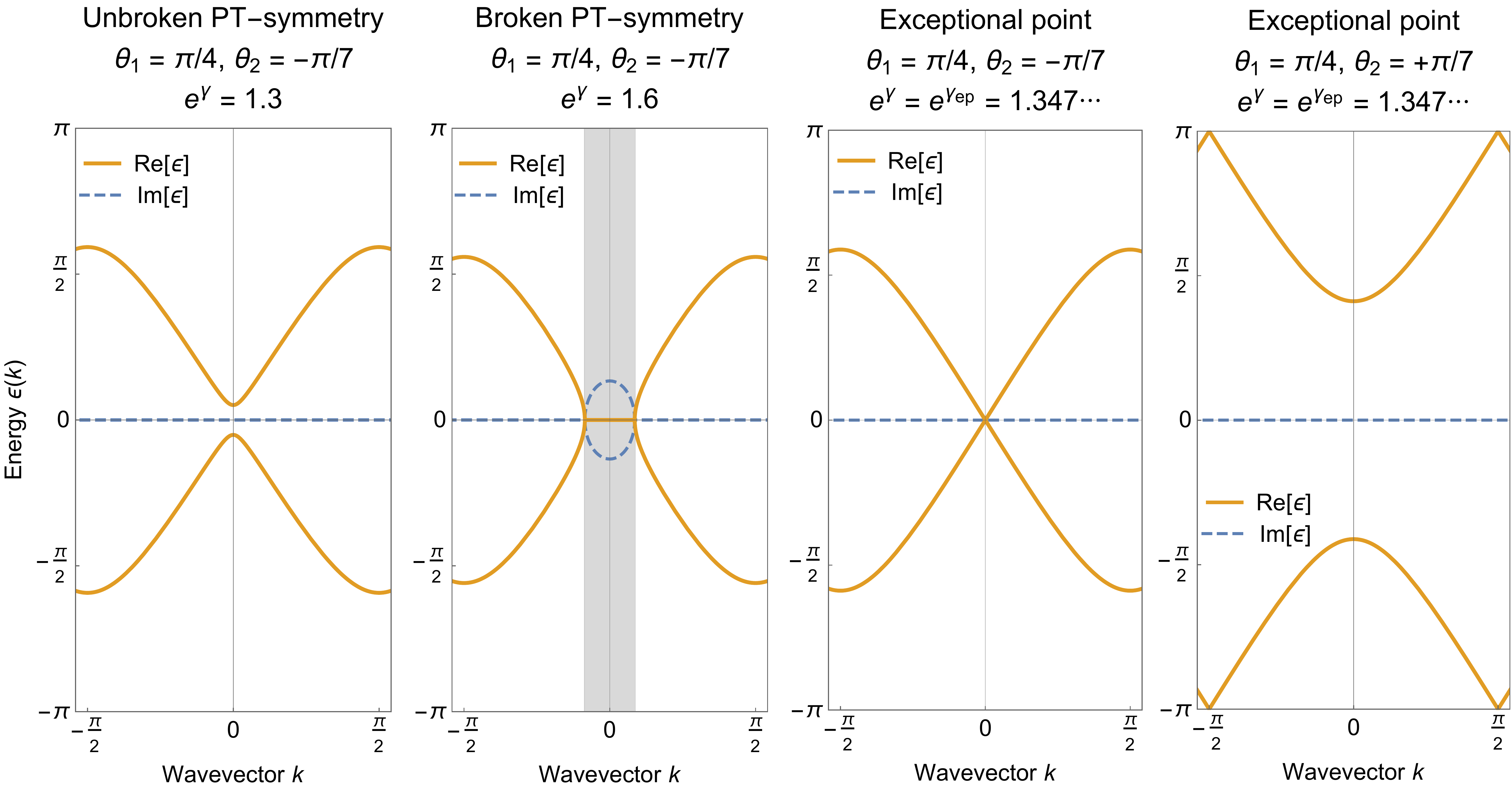}
\caption{In the unbroken symmetry phase of the model, the effective band structure is gapped  and real. In the broken PT-symmetry phase, the energy eigenvalues are purely imaginary over segments of wavevector space (shaded regions). At the exceptional points, level crossings at energies equal to 0 or $\pi$ occur. The solid and dashed lines indicate real and imaginary parts of the effective energies, respectively.} \label{fig:energy_spectrum}
\end{figure}

The symmetry of this quantum walk under a joint parity-time-reversal $\mathcal{PT}$ transformation has been established in detail \cite{mochizuki2016}. In summary, this symmetry is manifested by the fact that the evolution operator in a symmetry time frame $\tilde{U}_{\text{sym}}(k)\equiv\tilde{C}(\theta_1/2) \tilde{U}(k) \tilde{C}(-\theta_1/2)$ \cite{asboth2013} is transformed into its inverse under complex conjugation:
\begin{equation}
    \mathcal{PT}\tilde{U}_{\text{sym}}(k)\mathcal{PT}^{-1} = [\tilde{U}_{\text{sym}}(k)]^* = \tilde{U}_{\text{sym}}^{-1}(k).
\end{equation}
This symmetry leads to important topological consequences on the effective band structure of the model (Figure~\ref{fig:energy_spectrum}). The non-Hermitian effective Hamiltonian $\tilde{H}(k)$ defined through $\tilde{U}(k) = e^{-i \tilde{H}(k)}$ has two bands \cite{mochizuki2016}
\begin{equation}
 \epsilon_\pm(k) = \pm \cos^{-1}\Bigl[\cos \theta_1 \cos \theta_2 \cos (2k) - \sin \theta_1 \sin \theta_2 \cosh(2\gamma)\Bigr],
\end{equation}
and the unbroken PT-symmetry and broken PT-symmetry phases are readily identified by the appearance of imaginary eigenvalues in the latter \cite{bender1999}. At the exceptional point separating the two phases, {a degeneracy occurs at either $\epsilon = 0$ or at $\epsilon = \pi$. In addition to the eigenvalues becoming degenerate at this coalescence point, the eigenstates of the two bands actually become proportional to each other so that the mode evolution operator $\tilde{U}$ here fails to be diagonalizable.

\begin{figure}[t]
\centering
\includegraphics[width=0.48\linewidth]{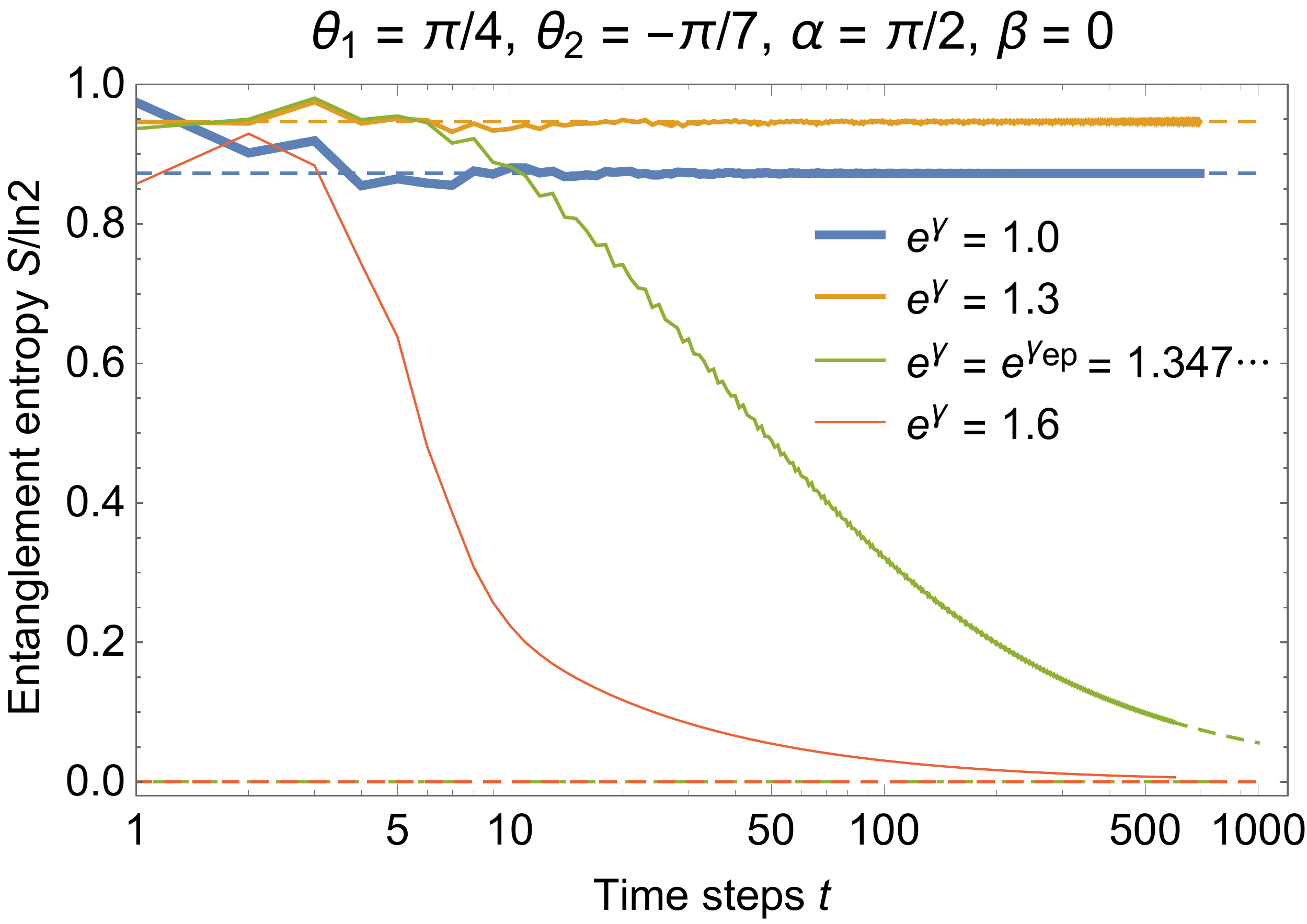}\includegraphics[width=0.5\linewidth]{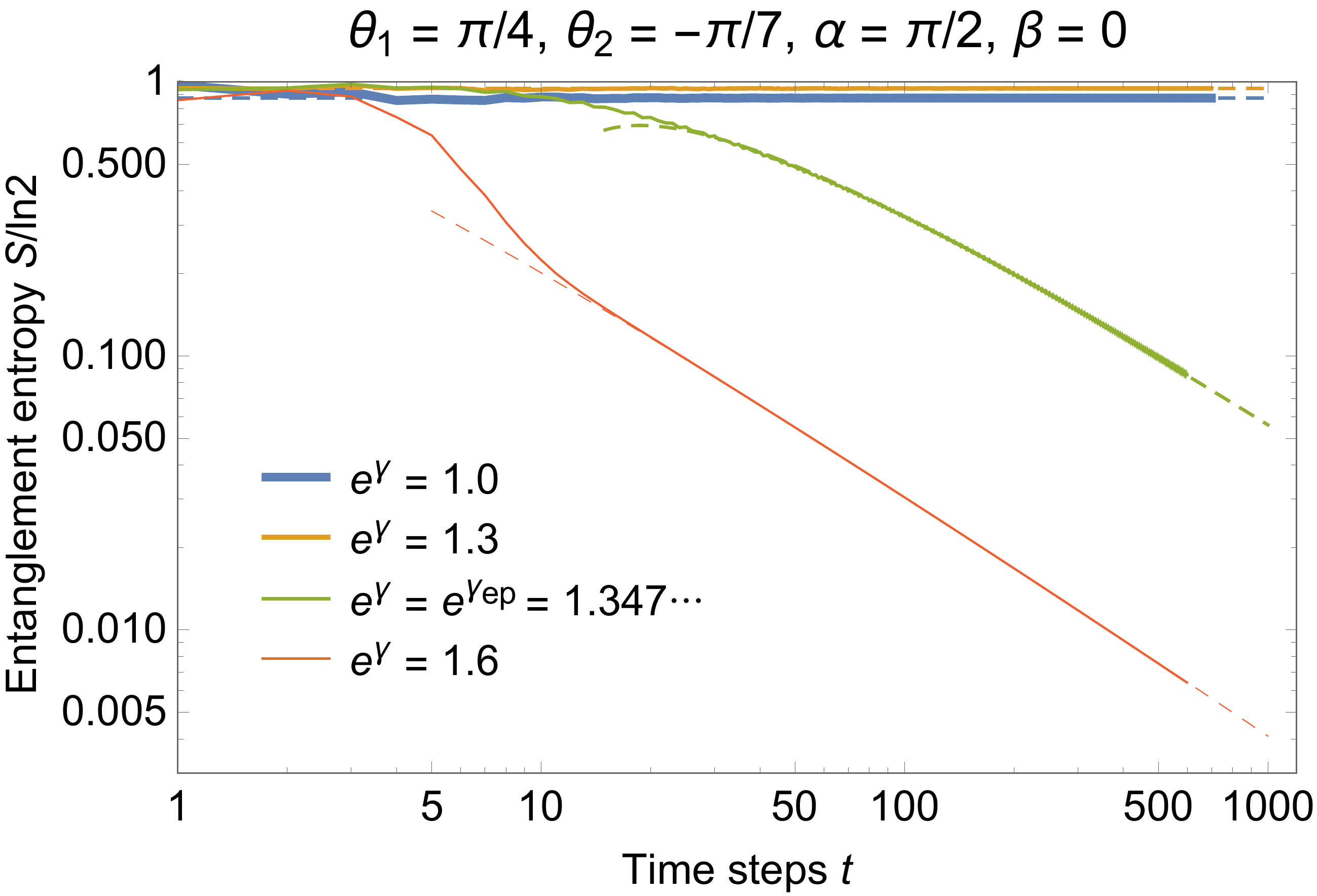}
\caption{Numerically exact calculations of the time evolution of the entanglement entropy reveal contrasting behaviors in different phases of the non-unitary quantum walk. In the unbroken PT-symmetry phase ($e^\gamma = 1.0$ and $e^\gamma = 1.3$) the entanglement entropy rapidly reaches a generally non-vanishing asymptotic value, which can be obtained by a stationary phase approach (dashed line). At the exceptional point  ($e^\gamma = e^{\gamma_\text{ep}}$) and in the broken PT-symmetry phase ($e^\gamma = 1.6$) the entanglement entropy eventually decays to zero. The double logarithmic plot on the right graphs the same data, but highlights the long-time decay (dashed lines) of the entanglement entropy in the broken PT-symmetry phase in Equation \eqref{eq:ptbdecay} and exceptional point in Equation \eqref{eq:epdecay}.} \label{fig:enttime}
\end{figure}

\section{Entanglement entropy for PT-symmetry phase diagram}\label{sec:entanglement}

In order to study the dynamics of hybrid entanglement in the model, we need to construct density operators at each time step of the quantum walk. However, the evolution is non-unitary and the conventional quantum mechanical operator $\ket{\psi(t)}\bra{\psi(t)}$ can not be interpreted as a statistical density operator. To remedy this problem, we adopt the approach of normalizing the state ket at every time step to get a properly normalized and Hermitian density operator
\begin{equation}
    \rho_\text{N}(t) = \frac{\ket{\psi(t)}\bra{\psi(t)}}{\trace\, \ket{\psi(t)}\bra{\psi(t)}}.
\end{equation}
This instantaneously normalized density operator $\rho_\text{N}(t)$ has been used to study entanglement in other non-Hermitian qudit models \cite{sergi2016quantum,wen2021} and fermionic chains \cite{herviou2019}, where it guarantees that reduced density operator spectra correspond to probabilities. Its use is further justified here because the raw intensity $\bra{\psi(t)}\psi(t)\rangle$ reproduces the observed amplified photon counts in the broken PT-symmetry phase of the corresponding experimental system \cite{regensburger2012parity,xiao2017,xiao2020}. Although there are other means of constructing density operators given a non-unitary evolution, such as a metric formulation of inner products that satisfies a number of quantum information 
no-go theorems \cite{ju2019,badhani2021non}, we have found that this  approach leads to a non-Hermitian reduced state with negative and sometimes complex entanglement entropy \cite{pechukas1994,herviou2019,chen2021} in an extensive region of our parameter space.

\begin{figure}[t]
\centering
\includegraphics[width=1\linewidth]{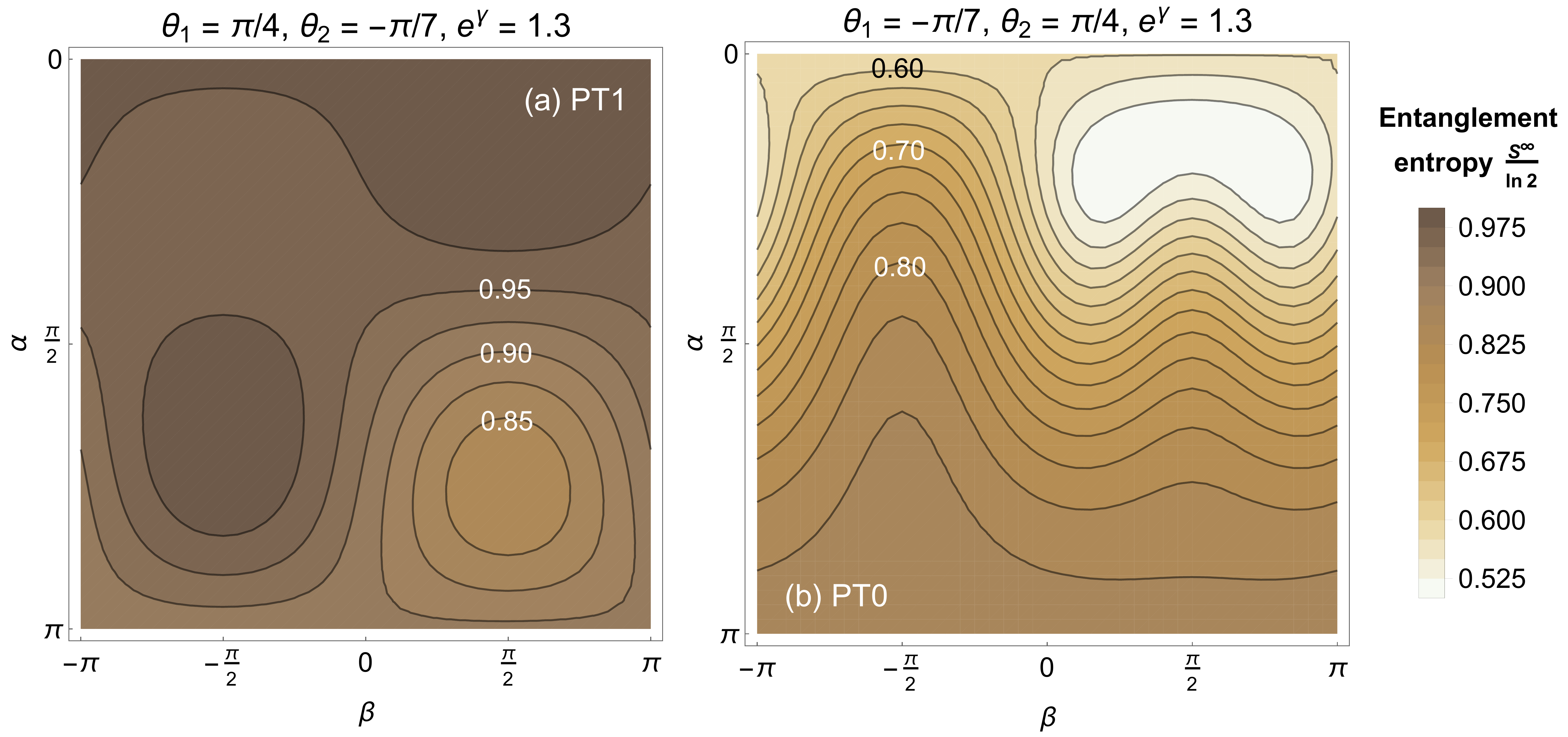}
\caption{There is persistent entanglement entropy $S^{\infty}$ in the unbroken PT-symmetry phase for arbitrary localized initial states $\ket{0}\otimes\cos{\alpha/2}\,\ket{L} + e^{i\beta}\sin{\alpha/2}\,\ket{R}$. The PT0 and PT1 parameter regions are shown on the phase diagrams in Figures \ref{fig:phase_diagram} and \ref{fig:winding}. }  \label{fig:initcond}
\end{figure}

For concreteness, we consider unentangled initial conditions that are localized at the origin $\ket{\psi(0)} = \ket{0}\otimes [\cos \alpha/2 \,\ket{L} + e^{i\beta}\sin\alpha/2 \,\ket{R}]$ (or $\ket{\tilde{\psi}_k(0)}  = \cos \alpha/2 \,\ket{L} + e^{i\beta}\sin\alpha/2 \,\ket{R}$ for all $k$), with $\alpha$ and $\beta$ polar and equatorial angles, respectively, on a Bloch sphere. 
 The reduced density operator on the coin space is then obtained as usual by partially tracing over the walker degrees of freedom $\rho_\text{c}(t) = \trace_\text{w} \,\ket{\psi_\text{N}(t)} \bra{\psi_\text{N}(t)}$ and the von Neumann entanglement entropy is $S(t)= -\trace [\rho_\text{c}(t) \ln \rho_\text{c}(t)]$.

 Since the evolution operator is partially diagonal in Fourier space, the numerical evaluation of the partial trace
 \begin{equation}\label{eq:rdm}
     \rho_\text{c}(t) = \frac{1}{\trace \rho_\text{c}(t)}\sum_{\sigma,\sigma'} \biggl[\int_{\text{BZ}}\, \iprod{\sigma}{\tilde{\psi}_k(t)}\iprod{\tilde{\psi}_k(t)}{\sigma'} \,\text{d}k \biggr]\ket{\sigma}\bra{\sigma'},
 \end{equation}
 is simple because the time-dependent kets $\ket{\tilde{\psi}_k(t)}=[\tilde{U}(k)]^t\ket{\tilde{\psi}_k(0)}$ can be solved in closed form by projection methods \cite{abal2006}. Furthermore, since the effective energies are either real or imaginary, the time dependence $e^{-i2\epsilon_\pm(k)t}$ of the terms in the integrand in Equation \ref{eq:rdm} allows one to obtain the long-time behavior of the entanglement entropy $S(t\to\infty) = S^{\infty}$ by a stationary phase approximation (real energies) or a steepest descent approach (imaginary energies).
 
 We use these results to analyze the dynamics of the hybrid entanglement between the coin and walker and find that we can easily distinguish between the PT phases of this non-unitary quantum walk (Figure~\ref{fig:enttime}). In the unbroken PT-symmetry phase, all of the eigenenergies are real so that the energy eigenstates evolve with phase angles that increase in time. Thus, a stationary phase analysis reveals that oscillations about the asymptotic stationary state eventually cancel out as the walk progresses and $S^\infty$ is generally non-zero (Figure~\ref{fig:initcond}).

The asymptotic entanglement entropy displays contrasting behavior when, instead, the PT-symmetry of the model is spontaneously broken by its energy eigenstates. In this case, hybrid entanglement that is produced in the first few steps of the walk decays to zero as the non-unitary walk progresses. Modes with imaginary effective energies have amplitudes that increase exponentially in time, so that an asymptotic analysis by a steepest descent method can be applied. In the infinite time limit $t\to \infty$ a product state is reached, which consists of the dominant walker mode with largest imaginary energy and a parameter-dependent coin state. Now, in the asymptotic expansions of the reduced density matrix elements resulting from the steepest descent method, the sub-leading term is $t^{-1}$ smaller than the leading term. Thus, we find that the entanglement spectrum (the eigenvalues of the reduced density operator)
also approaches that of a pure state as $t^{-1}$ (Figure~\ref{fig:enttime}) in the broken PT-symmetry phase:
\begin{equation}\label{eq:ptbdecay}
{S}(t\gg 1) \sim -\bigl[1-x(t)\bigr]\ln\bigl[1-x(t)\bigr] - x(t)\ln x(t), \text{\quad with } x(t) \propto t^{-1}.
\end{equation}

\begin{figure}[t]
\centering
\includegraphics[width=0.96\linewidth]{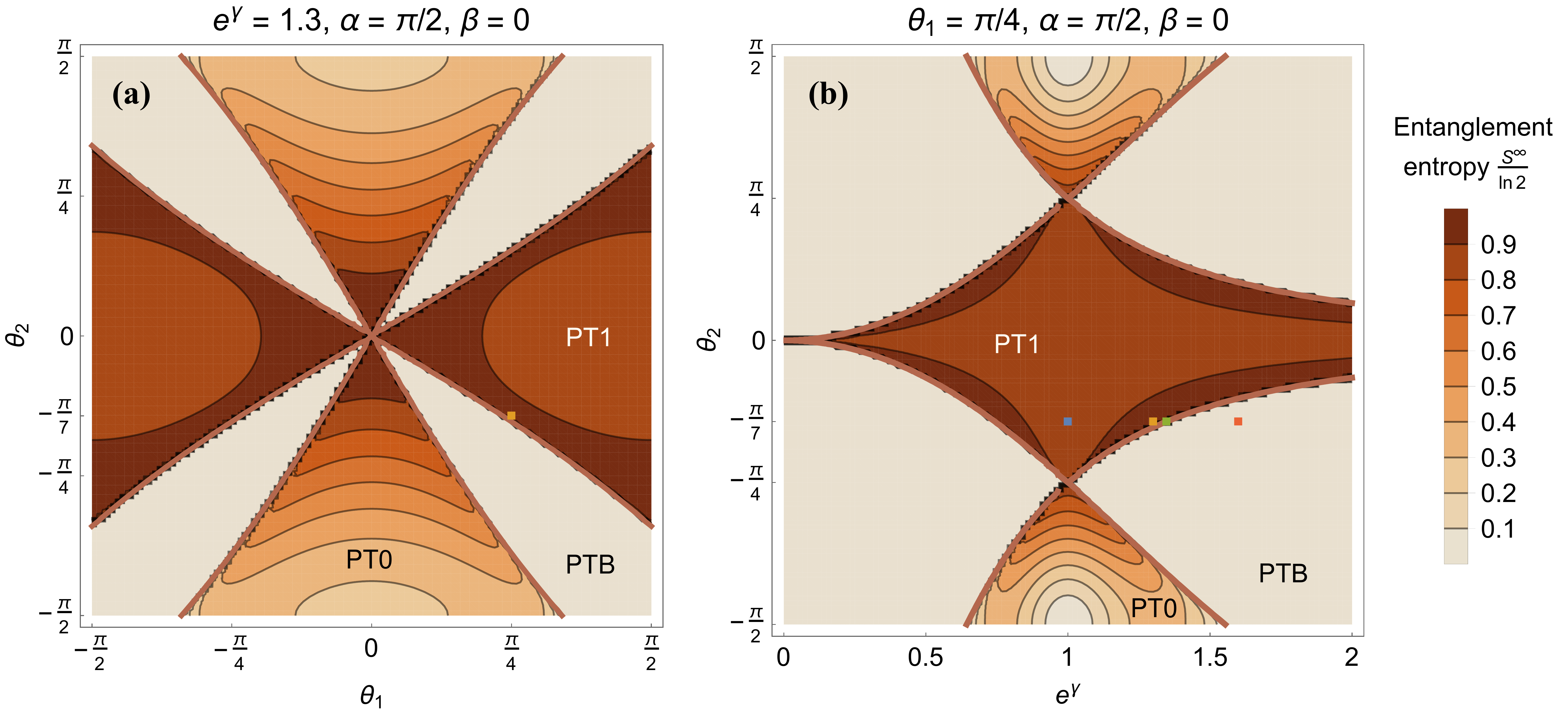}
\caption{In the non-unitary quantum walk model considered here, the asymptotic entanglement entropy ${S^\infty}$ vanishes in the broken PT-symmetry phase (PTB) and generally has non-zero values in the unbroken PT-symmetry phase (PT0 and PT1). The colored dots in these phase diagrams correspond to parameter values that were used to calculate the time-dependence of entanglement entropy in Figure \ref{fig:enttime}.} \label{fig:phase_diagram}
\end{figure}

At the exceptional point, a singular contribution to the density matrix elements comes from the level crossing at $k^* = 0$ when $\epsilon = 0$ (or $k^*=\pi/2$ when $\epsilon = \pi$), which dominates the long-time entanglement dynamics and leads to the eventual vanishing of entanglement. Since the evolution operator at the coalescence point $\tilde{U}(k^*)$ is not diagonalizable, a Jordan decomposition instead gives the time-dependence of the reduced density matrix elements as algebraically reaching their asymptotic values in time. The resulting entanglement entropy vanishes as
\begin{equation}\label{eq:epdecay}
{S}(t\gg 1) \sim -\bigl[1-x(t)\bigr]\ln\bigl[1-x(t)\bigr] - x(t)\ln x(t), \text{\quad with } x(t)= c_1 t^{-1} + c_2t^{-2},
\end{equation}
for constant, but parameter-dependent, $c_j$. We find that the decay to zero entanglement is much slower at the exceptional point than in the broken PT-symmetry phase (Figure \ref{fig:enttime}). For both cases, modes with real energies contribute to oscillating transient behavior. The reason for the slower decay at the exceptional point is that these transients are only suppressed polynomially in time by the singular contribution at the coalescence point. Meanwhile, in the broken PT-symmetry phase these transients are instead suppressed exponentially in time by a range of absorbing modes with imaginary energies.

\begin{figure}[t]
\centering
\includegraphics[width=0.96\linewidth]{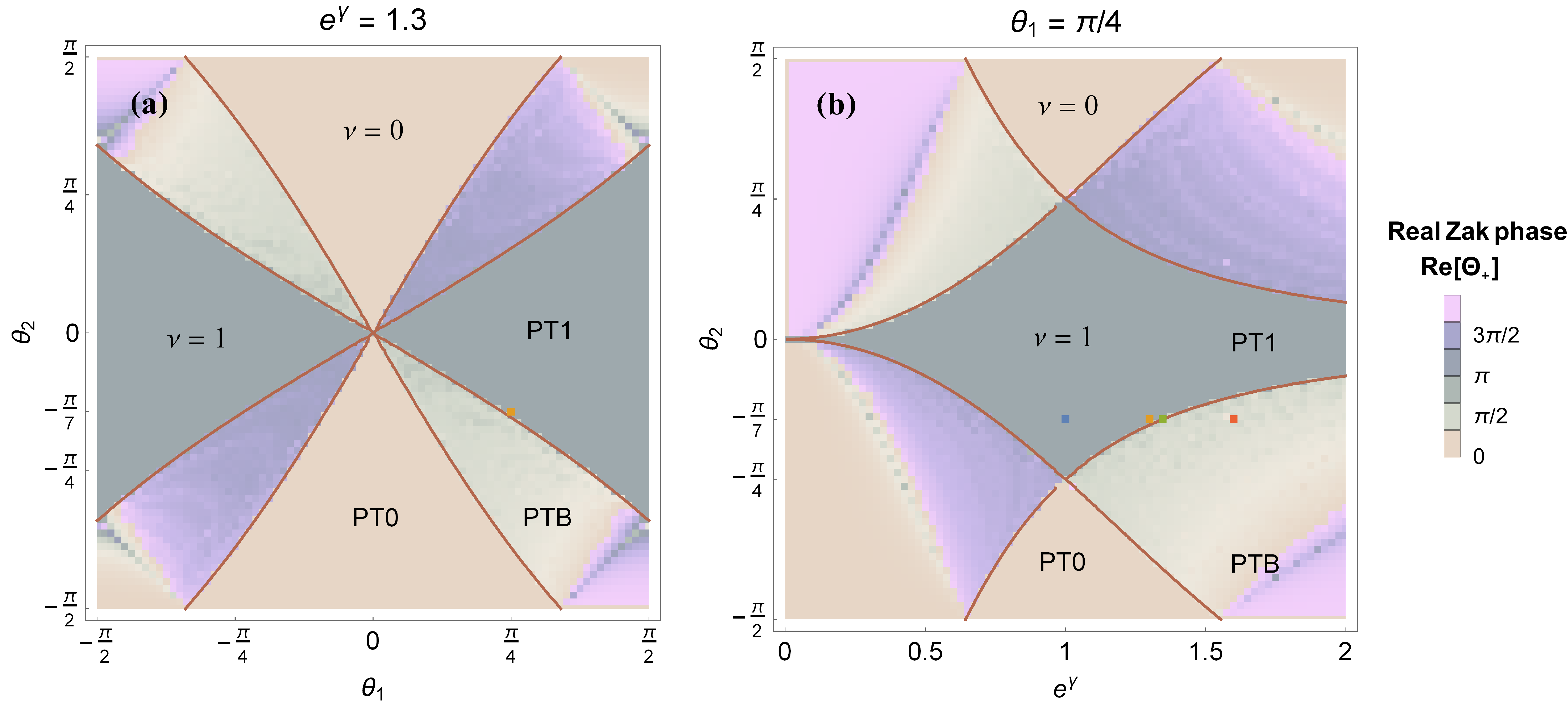}
\caption{The boundaries and differing features of the non-vanishing entanglement entropy seen in the PT-symmetry phase diagram (Figure \ref{fig:phase_diagram}) are also mirrored in the topological phase diagram of the model. A comparison reveals that the topological phase with winding number $\nu = 1$ (PT1) systematically has more entanglement than the topological phase with $\nu = 0$ (PT0). In the broken PT-symmetry phase (PTB), we plotted the real Zak phase modulo $2\pi$, and the color gradients indicate that $\nu$ is not quantized there.\label{fig:winding}} 
\end{figure}

These observations are summarized in phase diagrams (Figure \ref{fig:phase_diagram}) that delineate the unbroken and broken PT-symmetry phases of the model, where the asymptotic entanglement entropy $S^\infty$ is used as an indicator for the phase transition \cite{dechiara2012}. We find that critical lines consisting of exceptional points clearly separate the unbroken PT-symmetry phases (PT0 and PT1) with non-zero entanglement entropy from the broken PT-symmetry phase (PTB) with vanishing entanglement entropy.  

\section{Entanglement entropy can differentiate topological phases}\label{sec:topo}

Inspection of the phase diagrams (Figure \ref{fig:phase_diagram}) suggest that the unbroken PT-symmetry phase of the model may be divided further into two regions. About the multicritical points where two lines of exceptional points intersect, we see regions with slowly-varying asymptotic entanglement entropy (PT1) that can be distinguished from regions with systematically lower entanglement and larger gradients in the parameter space of the model (PT0).

In recent photonic experiments involving a similar split-step unitary quantum walk \cite{wang2020}, the entanglement entropy was found to be constant in the non-trivial  topological phase (non-zero winding number) of the system and decreased across a topological phase transition into the trivial topological phase (zero winding number). We observe the same behavior along the unitary plane $e^\gamma = 1$ (Figure \ref{fig:phase_diagram}b), which suggests that the entanglement entropy may also be used to identify bulk topological phases in this non-unitary quantum walk.

\begin{figure}[t]
\centering
\includegraphics[width=0.96\linewidth]{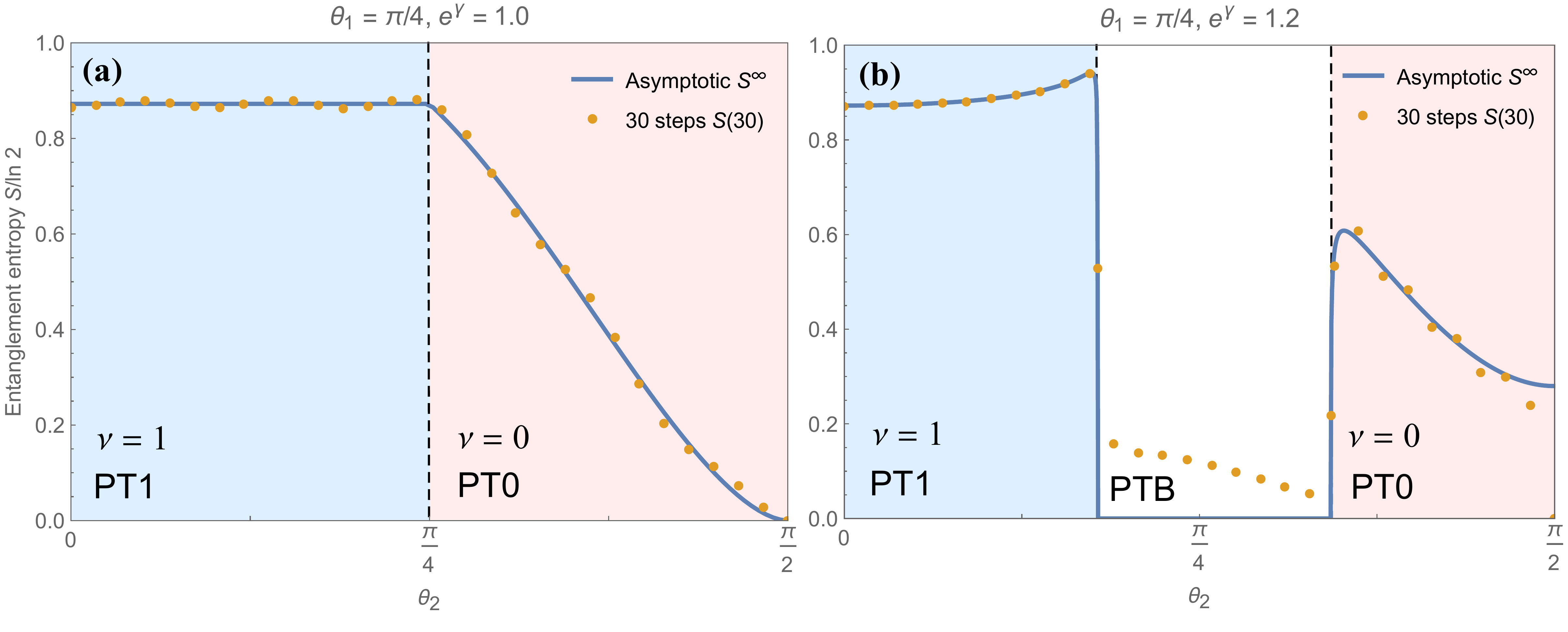}
\caption{(a) Along the unitary plane $e^\gamma = 1$, the unbroken PT-symmetry phases PT$\nu$ with integer winding number $\nu$ are adjacent to each other. (b) In the non-unitary case with $e^\gamma \ne 1$, the winding numbers of these phases are preserved, but non-topological gaps with broken PT-symmetry and vanishing entanglement now separate the topological phases. Asymptotic values of the entanglement entropy $S^\infty$ are depicted by solid lines and finite time results for 30 steps are shown as dots. \label{fig:gap}}
\end{figure}

However, the non-unitary character of the model prompts us to use a complex generalization of the Berry phase \cite{hatsugai2006,liang2013} for topological classification in the corresponding non-Hermitian Bloch system. That is, we define a generalized Zak phase \cite{zak1989} associated with the band $j$ as the Brillouin zone is traversed:
\begin{equation}\label{eq:zak}
    \Theta_j \equiv  \int_{-\pi/2}^{\pi/2} \bra{u_j^l}i\partial_k\ket{u_j^r}\,\text{d}k,
\end{equation}
where $\ket{u_j^r}$ and $\ket{u_j^l}$ are, respectively, the right and left eigenkets of the effective Hamiltonian $\tilde{H}(k)$ \cite{cardano2017,shen2018,zhang2019}. These eigenkets form a biorthogonal basis that satisfies the biorthonormality $\bra{u_j^{l}(k)}u_{j'}^r(k')\rangle = \delta_{jj'}\delta_{kk'}$ and completeness $\sum_{j}\ket{u_j^{r}(k)}\bra{u_{j}^l(k)} = \mathbb{1}_k$ relations \cite{brody2013}. A complete classification of the topological phases of this model has been presented previously \cite{xiao2017,mochizuki2020}, and it turns out that the real parts of the complex Zak phases are associated with integer winding numbers in the unbroken PT-symmetry phases:
\begin{equation}
    \nu_j \equiv \frac{1}{\pi}\real \Theta_j.
\end{equation}
In these topological phases the winding number is independent of band index ($\nu \equiv \nu_j$), but not so in the broken PT-symmetry phase. These results are summarized in the phase diagram Figure \ref{fig:winding}, where the quantization of the winding number $\nu$ across the unbroken PT-symmetry regions PT$\nu$ are seen. As observed in this and other non-unitary quantum walks with PT-symmetry breaking \cite{xiao2017,mochizuki2020,wang2022}, the quantized winding numbers are robust and unchanged as the system is driven away from unitarity, but only up to a certain threshold. When the strength of the gain-loss mechanism is increased beyond this limit and the exceptional points are crossed, there is a sudden loss of entanglement. 

Furthermore, we find that the phases in this topological phase diagram are indeed identified by features of the asymptotic entanglement illustrated in Figure \ref{fig:gap}: zero entanglement in the non-topological region (PTB), greater sensitivity to model parameters and overall lower entanglement in the $\nu = 0$ topological phase (PT0), and slowly-varying and higher entanglement in the $\nu = 1$ topological phase (PT1). The distinct behavior of the entanglement entropy in the PT0 and PT1 phases originate from the pole structure of the contour integral representation of the Zak phase \eqref{eq:zak}. At the exceptional points, poles of the analytically continued integrand fall on the contour, which leads to the entanglement entropy having different functional dependence on system parameters on either side of the phase transition \cite{wang2020}.

\section{Summary and outlook}\label{sec:conclusions}

We have demonstrated that the dynamics and asymptotic value of the coin-walker entanglement entropy can effectively differentiate the PT-symmetry and topological phases in a non-unitary quantum walk. Since this entanglement is determined by the reduced state, our result implies that these phases can be identified experimentally by measuring fluctuations \cite{song2012a} or performing full state tomography on just the qubit coin space. Doing so can complement current strategies that perform phase characterization and detection by measurements on the walker space \cite{regensburger2012parity,xiao2017,zhan2017,ozdemir2019}.

Another idea that could be pursued would be the possibility of engineering non-unitary evolutions or effective Hamiltonians to improve entanglement yields, for instance in the broken PT-symmetry phase. We have seen in the model here that entanglement vanished in asymptotic product states caused by the  dominating contribution of a single mode with imaginary energy. Perhaps some entanglement may be sustained in non-unitary quantum walks with sufficiently flat imaginary bands, or one with several peaks in the imaginary band structure.

This work also raises additional interesting  questions on the interactions between entanglement, non-unitarity, PT-symmetry, and topology. For example, noting that PT-symmetry exceptional points do not necessarily coincide with topological phase transitions \cite{acharya2022}, it would be worthwhile to study if the different characteristics of the persistent entanglement entropy we found are associated with PT-symmetry protection, topological order, or both. Also, the slow decay of entanglement at the exceptional point reminds us of critical behavior, and another possibly fruitful direction of study would be to investigate how the asymptotic entanglement entropy vanishes in more detail. PT-symmetry breaking is associated with a rather unique critical point \cite{ashida2017} and it may be possible to discover new universal dynamical scaling behavior in the entropy.

\section*{Acknowledgements}\label{sec:acknowledgements}

G.~M.~M.~Itable acknowledges support from the Department of Science and Technology--Science Education Institute through its Accelerated Science and Technology Human Resource Development Program. The authors are grateful for the guidance, mentorship, and valuable advice provided by J.~P.~H.~Esguerra.



\begin{thebibliography}{56}
\ifx \bisbn   \undefined \def \bisbn  #1{ISBN #1}\fi
\ifx \binits  \undefined \def \binits#1{#1}\fi
\ifx \bauthor  \undefined \def \bauthor#1{#1}\fi
\ifx \batitle  \undefined \def \batitle#1{#1}\fi
\ifx \bjtitle  \undefined \def \bjtitle#1{#1}\fi
\ifx \bvolume  \undefined \def \bvolume#1{\textbf{#1}}\fi
\ifx \byear  \undefined \def \byear#1{#1}\fi
\ifx \bissue  \undefined \def \bissue#1{#1}\fi
\ifx \bfpage  \undefined \def \bfpage#1{#1}\fi
\ifx \blpage  \undefined \def \blpage #1{#1}\fi
\ifx \burl  \undefined \def \burl#1{\textsf{#1}}\fi
\ifx \doiurl  \undefined \def \doiurl#1{\url{https://doi.org/#1}}\fi
\ifx \betal  \undefined \def \betal{\textit{et al.}}\fi
\ifx \binstitute  \undefined \def \binstitute#1{#1}\fi
\ifx \binstitutionaled  \undefined \def \binstitutionaled#1{#1}\fi
\ifx \bctitle  \undefined \def \bctitle#1{#1}\fi
\ifx \beditor  \undefined \def \beditor#1{#1}\fi
\ifx \bpublisher  \undefined \def \bpublisher#1{#1}\fi
\ifx \bbtitle  \undefined \def \bbtitle#1{#1}\fi
\ifx \bedition  \undefined \def \bedition#1{#1}\fi
\ifx \bseriesno  \undefined \def \bseriesno#1{#1}\fi
\ifx \blocation  \undefined \def \blocation#1{#1}\fi
\ifx \bsertitle  \undefined \def \bsertitle#1{#1}\fi
\ifx \bsnm \undefined \def \bsnm#1{#1}\fi
\ifx \bsuffix \undefined \def \bsuffix#1{#1}\fi
\ifx \bparticle \undefined \def \bparticle#1{#1}\fi
\ifx \barticle \undefined \def \barticle#1{#1}\fi
\ifx \bconfdate \undefined \def \bconfdate #1{#1}\fi
\ifx \botherref \undefined \def \botherref #1{#1}\fi
\ifx \url \undefined \def \url#1{\textsf{#1}}\fi
\ifx \bchapter \undefined \def \bchapter#1{#1}\fi
\ifx \bbook \undefined \def \bbook#1{#1}\fi
\ifx \bcomment \undefined \def \bcomment#1{#1}\fi
\ifx \oauthor \undefined \def \oauthor#1{#1}\fi
\ifx \citeauthoryear \undefined \def \citeauthoryear#1{#1}\fi
\ifx \endbibitem  \undefined \def \endbibitem {}\fi
\ifx \bconflocation  \undefined \def \bconflocation#1{#1}\fi
\ifx \arxivurl  \undefined \def \arxivurl#1{\textsf{#1}}\fi
\csname PreBibitemsHook\endcsname

\bibitem{nayak2000}
\begin{botherref}
\oauthor{\bsnm{Nayak}, \binits{A.}},
\oauthor{\bsnm{Vishwanath}, \binits{A.}}:
Quantum walk on the line.
e-print \href{arXiv:quant-ph/0010117}{https://arxiv.org/abs/quant-ph/0010117}
(2000)
\end{botherref}
\endbibitem

\bibitem{ambainis2003}
\begin{barticle}
\bauthor{\bsnm{Ambainis}, \binits{A.}}:
\batitle{Quantum walks and their algorithmic applications}.
\bjtitle{Int. J. Quantum Inform.}
\bvolume{1}(\bissue{4}),
\bfpage{507}--\blpage{518}
(\byear{2003}).
\doiurl{10.1142/S0219749903000383}
\end{barticle}
\endbibitem

\bibitem{childs2009universal}
\begin{barticle}
\bauthor{\bsnm{Childs}, \binits{A.M.}}:
\batitle{Universal computation by quantum walk}.
\bjtitle{Phys. Rev. Lett.}
\bvolume{102}(\bissue{18}),
\bfpage{180501}
(\byear{2009}).
\doiurl{10.1103/PhysRevLett.102.180501}
\end{barticle}
\endbibitem

\bibitem{lovett2010universal}
\begin{barticle}
\bauthor{\bsnm{Lovett}, \binits{N.B.}},
\bauthor{\bsnm{Cooper}, \binits{S.}},
\bauthor{\bsnm{Everitt}, \binits{M.}},
\bauthor{\bsnm{Trevers}, \binits{M.}},
\bauthor{\bsnm{Kendon}, \binits{V.}}:
\batitle{Universal quantum computation using the discrete-time quantum walk}.
\bjtitle{Phys. Rev. A}
\bvolume{81}(\bissue{4}),
\bfpage{042330}
(\byear{2010}).
\doiurl{10.1103/PhysRevA.81.042330}
\end{barticle}
\endbibitem

\bibitem{kurzy2011discrete}
\begin{barticle}
\bauthor{\bsnm{Kurzy\ifmmode\acute{n}\else\'{n}\fi{}ski}, \binits{P.}},
\bauthor{\bsnm{W\'ojcik}, \binits{A.}}:
\batitle{Discrete-time quantum walk approach to state transfer}.
\bjtitle{Phys. Rev. A}
\bvolume{83}(\bissue{6}),
\bfpage{062315}
(\byear{2011}).
\doiurl{10.1103/PhysRevA.83.062315}
\end{barticle}
\endbibitem

\bibitem{schreiber2010photons}
\begin{barticle}
\bauthor{\bsnm{Schreiber}, \binits{A.}},
\bauthor{\bsnm{Cassemiro}, \binits{K.N.}},
\bauthor{\bsnm{Poto{\v{c}}ek}, \binits{V.}},
\bauthor{\bsnm{G{\'a}bris}, \binits{A.}},
\bauthor{\bsnm{Mosley}, \binits{P.J.}},
\bauthor{\bsnm{Andersson}, \binits{E.}},
\bauthor{\bsnm{Jex}, \binits{I.}},
\bauthor{\bsnm{Silberhorn}, \binits{C.}}:
\batitle{Photons walking the line: a quantum walk with adjustable coin
  operations}.
\bjtitle{Phys. Rev. Lett.}
\bvolume{104}(\bissue{5}),
\bfpage{050502}
(\byear{2010}).
\doiurl{10.1103/PhysRevLett.104.050502}
\end{barticle}
\endbibitem

\bibitem{karski2009quantum}
\begin{barticle}
\bauthor{\bsnm{Karski}, \binits{M.}},
\bauthor{\bsnm{F{\"o}rster}, \binits{L.}},
\bauthor{\bsnm{Choi}, \binits{J.-M.}},
\bauthor{\bsnm{Steffen}, \binits{A.}},
\bauthor{\bsnm{Alt}, \binits{W.}},
\bauthor{\bsnm{Meschede}, \binits{D.}},
\bauthor{\bsnm{Widera}, \binits{A.}}:
\batitle{Quantum walk in position space with single optically trapped atoms}.
\bjtitle{Science}
\bvolume{325}(\bissue{5937}),
\bfpage{174}--\blpage{177}
(\byear{2009}).
\doiurl{10.1126/science.1174436}
\end{barticle}
\endbibitem

\bibitem{preiss2015}
\begin{barticle}
\bauthor{\bsnm{Preiss}, \binits{P.M.}},
\bauthor{\bsnm{Ma}, \binits{R.}},
\bauthor{\bsnm{Eric~Tai}, \binits{M.}},
\bauthor{\bsnm{Lukin}, \binits{A.}},
\bauthor{\bsnm{Rispoli}, \binits{M.}},
\bauthor{\bsnm{Zupancic}, \binits{P.}},
\bauthor{\bsnm{Lahini}, \binits{Y.}},
\bauthor{\bsnm{Islam}, \binits{R.}},
\bauthor{\bsnm{Greiner}, \binits{M.}}:
\batitle{Strongly correlated quantum walks in optical lattices}.
\bjtitle{Science}
\bvolume{347}(\bissue{6227}),
\bfpage{1229}--\blpage{1233}
(\byear{2015}).
\doiurl{10.1126/science.1260364}
\end{barticle}
\endbibitem

\bibitem{manouchehri2014}
\begin{bbook}
\bauthor{\bsnm{Manouchehri}, \binits{K.}},
\bauthor{\bsnm{Wang}, \binits{J.}}:
\bbtitle{Physical Implementation of Quantum Walks},
\bedition{1st} edn.
\bpublisher{Springer},
\blocation{Berlin}
(\byear{2014}).
\doiurl{10.1007/978-3-642-36014-5}
\end{bbook}
\endbibitem

\bibitem{denicola2014quantum}
\begin{barticle}
\bauthor{\bsnm{De~Nicola}, \binits{F.}},
\bauthor{\bsnm{Sansoni}, \binits{L.}},
\bauthor{\bsnm{Crespi}, \binits{A.}},
\bauthor{\bsnm{Ramponi}, \binits{R.}},
\bauthor{\bsnm{Osellame}, \binits{R.}},
\bauthor{\bsnm{Giovannetti}, \binits{V.}},
\bauthor{\bsnm{Fazio}, \binits{R.}},
\bauthor{\bsnm{Mataloni}, \binits{P.}},
\bauthor{\bsnm{Sciarrino}, \binits{F.}}:
\batitle{Quantum simulation of bosonic-fermionic noninteracting particles in
  disordered systems via a quantum walk}.
\bjtitle{Phys. Rev. A}
\bvolume{89}(\bissue{3}),
\bfpage{032322}
(\byear{2014}).
\doiurl{10.1103/PhysRevA.89.032322}
\end{barticle}
\endbibitem

\bibitem{regensburger2012parity}
\begin{barticle}
\bauthor{\bsnm{Regensburger}, \binits{A.}},
\bauthor{\bsnm{Bersch}, \binits{C.}},
\bauthor{\bsnm{Miri}, \binits{M.-A.}},
\bauthor{\bsnm{Onishchukov}, \binits{G.}},
\bauthor{\bsnm{Christodoulides}, \binits{D.N.}},
\bauthor{\bsnm{Peschel}, \binits{U.}}:
\batitle{Parity--time synthetic photonic lattices}.
\bjtitle{Nature}
\bvolume{488}(\bissue{7410}),
\bfpage{167}--\blpage{171}
(\byear{2012}).
\doiurl{10.1038/nature11298}
\end{barticle}
\endbibitem

\bibitem{xiao2017}
\begin{barticle}
\bauthor{\bsnm{Xiao}, \binits{L.}},
\bauthor{\bsnm{Zhan}, \binits{X.}},
\bauthor{\bsnm{Bian}, \binits{Z.}},
\bauthor{\bsnm{Wang}, \binits{K.}},
\bauthor{\bsnm{Zhang}, \binits{X.}},
\bauthor{\bsnm{Wang}, \binits{X.}},
\bauthor{\bsnm{Li}, \binits{J.}},
\bauthor{\bsnm{Mochizuki}, \binits{K.}},
\bauthor{\bsnm{Kim}, \binits{D.}},
\bauthor{\bsnm{Kawakami}, \binits{N.}},
\bauthor{\bsnm{Yi}, \binits{W.}},
\bauthor{\bsnm{Obuse}, \binits{H.}},
\bauthor{\bsnm{Sanders}, \binits{B.C.}},
\bauthor{\bsnm{Xue}, \binits{P.}}:
\batitle{Observation of topological edge states in parity--time-symmetric
  quantum walks}.
\bjtitle{Nat. Phys.}
\bvolume{13}(\bissue{11}),
\bfpage{1117}--\blpage{1123}
(\byear{2017}).
\doiurl{10.1038/nphys4204}
\end{barticle}
\endbibitem

\bibitem{zhan2017}
\begin{barticle}
\bauthor{\bsnm{Zhan}, \binits{X.}},
\bauthor{\bsnm{Xiao}, \binits{L.}},
\bauthor{\bsnm{Bian}, \binits{Z.}},
\bauthor{\bsnm{Wang}, \binits{K.}},
\bauthor{\bsnm{Qiu}, \binits{X.}},
\bauthor{\bsnm{Sanders}, \binits{B.C.}},
\bauthor{\bsnm{Yi}, \binits{W.}},
\bauthor{\bsnm{Xue}, \binits{P.}}:
\batitle{Detecting topological invariants in nonunitary discrete-time quantum
  walks}.
\bjtitle{Phys. Rev. Lett.}
\bvolume{119}(\bissue{13}),
\bfpage{130501}
(\byear{2017}).
\doiurl{10.1103/PhysRevLett.119.130501}
\end{barticle}
\endbibitem

\bibitem{ozdemir2019}
\begin{barticle}
\bauthor{\bsnm{{\"O}zdemir}, \binits{{\c{S}}.K.}},
\bauthor{\bsnm{Rotter}, \binits{S.}},
\bauthor{\bsnm{Nori}, \binits{F.}},
\bauthor{\bsnm{Yang}, \binits{L.}}:
\batitle{Parity--time symmetry and exceptional points in photonics}.
\bjtitle{Nat. Mater.}
\bvolume{18}(\bissue{8}),
\bfpage{783}--\blpage{798}
(\byear{2019}).
\doiurl{10.1038/s41563-019-0304-9}
\end{barticle}
\endbibitem

\bibitem{hatano2021}
\begin{barticle}
\bauthor{\bsnm{Hatano}, \binits{N.}},
\bauthor{\bsnm{Obuse}, \binits{H.}}:
\batitle{Delocalization of a non-{H}ermitian quantum walk on random media in
  one dimension}.
\bjtitle{Ann. Phys. (NY)}
\bvolume{435},
\bfpage{168615}
(\byear{2021}).
\doiurl{10.1016/j.aop.2021.168615}
\end{barticle}
\endbibitem

\bibitem{lin2022}
\begin{barticle}
\bauthor{\bsnm{Lin}, \binits{Q.}},
\bauthor{\bsnm{Li}, \binits{T.}},
\bauthor{\bsnm{Xiao}, \binits{L.}},
\bauthor{\bsnm{Wang}, \binits{K.}},
\bauthor{\bsnm{Yi}, \binits{W.}},
\bauthor{\bsnm{Xue}, \binits{P.}}:
\batitle{Observation of non-{H}ermitian topological {A}nderson insulator in
  quantum dynamics}.
\bjtitle{Nat. Commun.}
\bvolume{13}(\bissue{1}),
\bfpage{3229}
(\byear{2022}).
\doiurl{10.1038/s41467-022-30938-9}
\end{barticle}
\endbibitem

\bibitem{carneiro2005}
\begin{barticle}
\bauthor{\bsnm{Carneiro}, \binits{I.}},
\bauthor{\bsnm{Loo}, \binits{M.}},
\bauthor{\bsnm{Xu}, \binits{X.}},
\bauthor{\bsnm{Girerd}, \binits{M.}},
\bauthor{\bsnm{Kendon}, \binits{V.}},
\bauthor{\bsnm{Knight}, \binits{P.L.}}:
\batitle{Entanglement in coined quantum walks on regular graphs}.
\bjtitle{New J. Phys.}
\bvolume{7}(\bissue{1}),
\bfpage{156}
(\byear{2005}).
\doiurl{10.1088/1367-2630/7/1/156}
\end{barticle}
\endbibitem

\bibitem{abal2006}
\begin{barticle}
\bauthor{\bsnm{Abal}, \binits{G.}},
\bauthor{\bsnm{Siri}, \binits{R.}},
\bauthor{\bsnm{Romanelli}, \binits{A.}},
\bauthor{\bsnm{Donangelo}, \binits{R.}}:
\batitle{Quantum walk on the line: {E}ntanglement and nonlocal initial
  conditions}.
\bjtitle{Phys. Rev. A}
\bvolume{73}(\bissue{4}),
\bfpage{042302}
(\byear{2006}).
\doiurl{10.1103/PhysRevA.73.042302}
\end{barticle}
\endbibitem

\bibitem{ide2011}
\begin{barticle}
\bauthor{\bsnm{Ide}, \binits{Y.}},
\bauthor{\bsnm{Konno}, \binits{N.}},
\bauthor{\bsnm{Machida}, \binits{T.}}:
\batitle{Entanglement for discrete-time quantum walks on the line}.
\bjtitle{Quantum Inf. Comput.}
\bvolume{11}(\bissue{9-10}),
\bfpage{855}--\blpage{866}
(\byear{2011}).
\bcomment{e-print
  \href{arXiv:1012.4164}{https://doi.org/10.48550/arXiv.1012.4164}}
\end{barticle}
\endbibitem

\bibitem{neves2009hybrid}
\begin{barticle}
\bauthor{\bsnm{Neves}, \binits{L.}},
\bauthor{\bsnm{Lima}, \binits{G.}},
\bauthor{\bsnm{Delgado}, \binits{A.}},
\bauthor{\bsnm{Saavedra}, \binits{C.}}:
\batitle{Hybrid photonic entanglement: Realization, characterization, and
  applications}.
\bjtitle{Phys. Rev. A}
\bvolume{80}(\bissue{4}),
\bfpage{042322}
(\byear{2009}).
\doiurl{10.1103/PhysRevA.80.042322}
\end{barticle}
\endbibitem

\bibitem{li2018hyper}
\begin{barticle}
\bauthor{\bsnm{Li}, \binits{Y.}},
\bauthor{\bsnm{Gessner}, \binits{M.}},
\bauthor{\bsnm{Li}, \binits{W.}},
\bauthor{\bsnm{Smerzi}, \binits{A.}}:
\batitle{Hyper- and hybrid nonlocality}.
\bjtitle{Phys. Rev. Lett.}
\bvolume{120},
\bfpage{050404}
(\byear{2018}).
\doiurl{10.1103/PhysRevLett.120.050404}
\end{barticle}
\endbibitem

\bibitem{flamini2018photonic}
\begin{barticle}
\bauthor{\bsnm{Flamini}, \binits{F.}},
\bauthor{\bsnm{Spagnolo}, \binits{N.}},
\bauthor{\bsnm{Sciarrino}, \binits{F.}}:
\batitle{Photonic quantum information processing: a review}.
\bjtitle{Rep. Prog. Phys.}
\bvolume{82}(\bissue{1}),
\bfpage{016001}
(\byear{2019}).
\doiurl{10.1088/1361-6633/aad5b2}
\end{barticle}
\endbibitem

\bibitem{gratsea2020}
\begin{barticle}
\bauthor{\bsnm{Gratsea}, \binits{A.}},
\bauthor{\bsnm{Metz}, \binits{F.}},
\bauthor{\bsnm{Busch}, \binits{T.}}:
\batitle{Universal and optimal coin sequences for high entanglement generation
  in 1{D} discrete time quantum walks}.
\bjtitle{J. Phys. A: Math. Theor.}
\bvolume{53}(\bissue{44}),
\bfpage{445306}
(\byear{2020}).
\doiurl{10.1088/1751-8121/abb54d}
\end{barticle}
\endbibitem

\bibitem{gratsea2020b}
\begin{barticle}
\bauthor{\bsnm{Gratsea}, \binits{A.}},
\bauthor{\bsnm{Lewenstein}, \binits{M.}},
\bauthor{\bsnm{Dauphin}, \binits{A.}}:
\batitle{Generation of hybrid maximally entangled states in a one-dimensional
  quantum walk}.
\bjtitle{Quantum Sci. Technol.}
\bvolume{5}(\bissue{2}),
\bfpage{025002}
(\byear{2020}).
\doiurl{10.1088/2058-9565/ab6ce6}
\end{barticle}
\endbibitem

\bibitem{vieira2013}
\begin{barticle}
\bauthor{\bsnm{Vieira}, \binits{R.}},
\bauthor{\bsnm{Amorim}, \binits{E.P.M.}},
\bauthor{\bsnm{Rigolin}, \binits{G.}}:
\batitle{Dynamically disordered quantum walk as a maximal entanglement
  generator}.
\bjtitle{Phys. Rev. Lett.}
\bvolume{111}(\bissue{18}),
\bfpage{180503}
(\byear{2013}).
\doiurl{10.1103/PhysRevLett.111.180503}
\end{barticle}
\endbibitem

\bibitem{maloyer2007}
\begin{barticle}
\bauthor{\bsnm{Maloyer}, \binits{O.}},
\bauthor{\bsnm{Kendon}, \binits{V.}}:
\batitle{Decoherence versus entanglement in coined quantum walks}.
\bjtitle{New J. Phys.}
\bvolume{9}(\bissue{4}),
\bfpage{87}
(\byear{2007}).
\doiurl{10.1088/1367-2630/9/4/087}
\end{barticle}
\endbibitem

\bibitem{dey2019}
\begin{barticle}
\bauthor{\bsnm{Dey}, \binits{S.}},
\bauthor{\bsnm{Raj}, \binits{A.}},
\bauthor{\bsnm{Goyal}, \binits{S.K.}}:
\batitle{Controlling decoherence via {PT}-symmetric non-{H}ermitian open
  quantum systems}.
\bjtitle{Phys. Lett. A}
\bvolume{383}(\bissue{30}),
\bfpage{125931}
(\byear{2019}).
\doiurl{10.1016/j.physleta.2019.125931}
\end{barticle}
\endbibitem

\bibitem{fring2019}
\begin{barticle}
\bauthor{\bsnm{Fring}, \binits{A.}},
\bauthor{\bsnm{Frith}, \binits{T.}}:
\batitle{Eternal life of entropy in non-{H}ermitian quantum systems}.
\bjtitle{Phys. Rev. A}
\bvolume{100}(\bissue{1}),
\bfpage{010102}
(\byear{2019}).
\doiurl{10.1103/PhysRevA.100.010102}
\end{barticle}
\endbibitem

\bibitem{chakraborty2019delayed}
\begin{barticle}
\bauthor{\bsnm{Chakraborty}, \binits{S.}},
\bauthor{\bsnm{Sarma}, \binits{A.K.}}:
\batitle{Delayed sudden death of entanglement at exceptional points}.
\bjtitle{Phys. Rev. A}
\bvolume{100}(\bissue{6}),
\bfpage{063846}
(\byear{2019}).
\doiurl{10.1103/PhysRevA.100.063846}
\end{barticle}
\endbibitem

\bibitem{bender1998a}
\begin{barticle}
\bauthor{\bsnm{Bender}, \binits{C.M.}},
\bauthor{\bsnm{Boettcher}, \binits{S.}}:
\batitle{Real spectra in non-{H}ermitian {H}amiltonians having
  $\mathcal{P}\mathcal{T}$ symmetry}.
\bjtitle{Phys. Rev. Lett.}
\bvolume{80}(\bissue{24}),
\bfpage{5243}--\blpage{5246}
(\byear{1998}).
\doiurl{10.1103/PhysRevLett.80.5243}
\end{barticle}
\endbibitem

\bibitem{bender1999}
\begin{barticle}
\bauthor{\bsnm{Bender}, \binits{C.M.}},
\bauthor{\bsnm{Boettcher}, \binits{S.}},
\bauthor{\bsnm{Meisinger}, \binits{P.N.}}:
\batitle{$\mathcal{P}\mathcal{T}$-symmetric quantum mechanics}.
\bjtitle{J. Math. Phys.}
\bvolume{40}(\bissue{5}),
\bfpage{2201}--\blpage{2229}
(\byear{1999}).
\doiurl{10.1063/1.532860}
\end{barticle}
\endbibitem

\bibitem{mochizuki2016}
\begin{barticle}
\bauthor{\bsnm{Mochizuki}, \binits{K.}},
\bauthor{\bsnm{Kim}, \binits{D.}},
\bauthor{\bsnm{Obuse}, \binits{H.}}:
\batitle{Explicit definition of $\mathscr{PT}$ symmetry for nonunitary quantum
  walks with gain and loss}.
\bjtitle{Phys. Rev. A}
\bvolume{93}(\bissue{6}),
\bfpage{062116}
(\byear{2016}).
\doiurl{10.1103/PhysRevA.93.062116}
\end{barticle}
\endbibitem

\bibitem{lambert2005}
\begin{barticle}
\bauthor{\bsnm{Lambert}, \binits{N.}},
\bauthor{\bsnm{Emary}, \binits{C.}},
\bauthor{\bsnm{Brandes}, \binits{T.}}:
\batitle{Entanglement and entropy in a spin-boson quantum phase transition}.
\bjtitle{Phys. Rev. A}
\bvolume{71}(\bissue{5}),
\bfpage{053804}
(\byear{2005}).
\doiurl{10.1103/PhysRevA.71.053804}
\end{barticle}
\endbibitem

\bibitem{dechiara2012}
\begin{barticle}
\bauthor{\bsnm{De~Chiara}, \binits{G.}},
\bauthor{\bsnm{Lepori}, \binits{L.}},
\bauthor{\bsnm{Lewenstein}, \binits{M.}},
\bauthor{\bsnm{Sanpera}, \binits{A.}}:
\batitle{Entanglement spectrum, critical exponents, and order parameters in
  quantum spin chains}.
\bjtitle{Phys. Rev. Lett.}
\bvolume{109}(\bissue{23}),
\bfpage{237208}
(\byear{2012}).
\doiurl{10.1103/PhysRevLett.109.237208}
\end{barticle}
\endbibitem

\bibitem{wang2020}
\begin{barticle}
\bauthor{\bsnm{Wang}, \binits{Q.-Q.}},
\bauthor{\bsnm{Xu}, \binits{X.-Y.}},
\bauthor{\bsnm{Pan}, \binits{W.-W.}},
\bauthor{\bsnm{Tao}, \binits{S.-J.}},
\bauthor{\bsnm{Chen}, \binits{Z.}},
\bauthor{\bsnm{Zhan}, \binits{Y.-T.}},
\bauthor{\bsnm{Sun}, \binits{K.}},
\bauthor{\bsnm{Xu}, \binits{J.-S.}},
\bauthor{\bsnm{Chen}, \binits{G.}},
\bauthor{\bsnm{Han}, \binits{Y.-J.}},
\bauthor{\bsnm{Li}, \binits{C.-F.}},
\bauthor{\bsnm{Guo}, \binits{G.-C.}}:
\batitle{Robustness of entanglement as an indicator of topological phases in
  quantum walks}.
\bjtitle{Optica}
\bvolume{7}(\bissue{1}),
\bfpage{53}--\blpage{58}
(\byear{2020}).
\doiurl{10.1364/OPTICA.375388}
\end{barticle}
\endbibitem

\bibitem{mochizuki2020}
\begin{barticle}
\bauthor{\bsnm{Mochizuki}, \binits{K.}},
\bauthor{\bsnm{Kim}, \binits{D.}},
\bauthor{\bsnm{Kawakami}, \binits{N.}},
\bauthor{\bsnm{Obuse}, \binits{H.}}:
\batitle{Bulk-edge correspondence in nonunitary {F}loquet systems with chiral
  symmetry}.
\bjtitle{Phys. Rev. A}
\bvolume{102}(\bissue{6}),
\bfpage{062202}
(\byear{2020}).
\doiurl{10.1103/PhysRevA.102.062202}
\end{barticle}
\endbibitem

\bibitem{wang2022}
\begin{barticle}
\bauthor{\bsnm{Wang}, \binits{Q.}},
\bauthor{\bsnm{Li}, \binits{Z.-J.}}:
\batitle{Topological invariants of nonunitary quantum walk with chiral
  symmetry}.
\bjtitle{Results Phys.}
\bvolume{34},
\bfpage{105279}
(\byear{2022}).
\doiurl{10.1016/j.rinp.2022.105279}
\end{barticle}
\endbibitem

\bibitem{asboth2013}
\begin{barticle}
\bauthor{\bsnm{Asb\'oth}, \binits{J.K.}},
\bauthor{\bsnm{Obuse}, \binits{H.}}:
\batitle{Bulk-boundary correspondence for chiral symmetric quantum walks}.
\bjtitle{Phys. Rev. B}
\bvolume{88}(\bissue{12}),
\bfpage{121406}
(\byear{2013}).
\doiurl{10.1103/PhysRevB.88.121406}
\end{barticle}
\endbibitem

\bibitem{sergi2016quantum}
\begin{barticle}
\bauthor{\bsnm{Sergi}, \binits{A.}},
\bauthor{\bsnm{Zloshchastiev}, \binits{K.G.}}:
\batitle{Quantum entropy of systems described by non-{H}ermitian
  {H}amiltonians}.
\bjtitle{J. Stat. Mech.: Theor. Exp.}
\bvolume{2016}(\bissue{3}),
\bfpage{033102}
(\byear{2016}).
\doiurl{10.1088/1742-5468/2016/03/033102}
\end{barticle}
\endbibitem

\bibitem{wen2021}
\begin{barticle}
\bauthor{\bsnm{Wen}, \binits{J.}},
\bauthor{\bsnm{Zheng}, \binits{C.}},
\bauthor{\bsnm{Ye}, \binits{Z.}},
\bauthor{\bsnm{Xin}, \binits{T.}},
\bauthor{\bsnm{Long}, \binits{G.}}:
\batitle{Stable states with nonzero entropy under broken $\mathcal{PT}$
  symmetry}.
\bjtitle{Phys. Rev. Research}
\bvolume{3}(\bissue{1}),
\bfpage{013256}
(\byear{2021}).
\doiurl{10.1103/PhysRevResearch.3.013256}
\end{barticle}
\endbibitem

\bibitem{herviou2019}
\begin{barticle}
\bauthor{\bsnm{Herviou}, \binits{L.}},
\bauthor{\bsnm{Regnault}, \binits{N.}},
\bauthor{\bsnm{Bardarson}, \binits{J.H.}}:
\batitle{{Entanglement spectrum and symmetries in non-{H}ermitian fermionic
  non-interacting models}}.
\bjtitle{SciPost Phys.}
\bvolume{7},
\bfpage{069}
(\byear{2019}).
\doiurl{10.21468/SciPostPhys.7.5.069}
\end{barticle}
\endbibitem

\bibitem{xiao2020}
\begin{barticle}
\bauthor{\bsnm{Xiao}, \binits{L.}},
\bauthor{\bsnm{Deng}, \binits{T.}},
\bauthor{\bsnm{Wang}, \binits{K.}},
\bauthor{\bsnm{Zhu}, \binits{G.}},
\bauthor{\bsnm{Wang}, \binits{Z.}},
\bauthor{\bsnm{Yi}, \binits{W.}},
\bauthor{\bsnm{Xue}, \binits{P.}}:
\batitle{Non-{H}ermitian bulk--boundary correspondence in quantum dynamics}.
\bjtitle{Nature Phys.}
\bvolume{16}(\bissue{7}),
\bfpage{761}--\blpage{766}
(\byear{2020}).
\doiurl{10.1038/s41567-020-0836-6}
\end{barticle}
\endbibitem

\bibitem{ju2019}
\begin{barticle}
\bauthor{\bsnm{Ju}, \binits{C.-Y.}},
\bauthor{\bsnm{Miranowicz}, \binits{A.}},
\bauthor{\bsnm{Chen}, \binits{G.-Y.}},
\bauthor{\bsnm{Nori}, \binits{F.}}:
\batitle{Non-{H}ermitian {H}amiltonians and no-go theorems in quantum
  information}.
\bjtitle{Phys. Rev. A}
\bvolume{100}(\bissue{6}),
\bfpage{062118}
(\byear{2019}).
\doiurl{10.1103/PhysRevA.100.062118}
\end{barticle}
\endbibitem

\bibitem{badhani2021non}
\begin{botherref}
\oauthor{\bsnm{Badhani}, \binits{H.}},
\oauthor{\bsnm{Banerjee}, \binits{S.}},
\oauthor{\bsnm{Chandrashekar}, \binits{C.}}:
Non-{H}ermitian quantum walks and non-Markovianity: the {coin-position}
  interaction.
e-print \href{arXiv:2109.10682}{https://arxiv.org/abs/2109.10682}
(2021)
\end{botherref}
\endbibitem

\bibitem{pechukas1994}
\begin{barticle}
\bauthor{\bsnm{Pechukas}, \binits{P.}}:
\batitle{Reduced dynamics need not be completely positive}.
\bjtitle{Phys. Rev. Lett.}
\bvolume{73}(\bissue{8}),
\bfpage{1060}--\blpage{1062}
(\byear{1994}).
\doiurl{10.1103/PhysRevLett.73.1060}
\end{barticle}
\endbibitem

\bibitem{chen2021}
\begin{barticle}
\bauthor{\bsnm{Chen}, \binits{L.-M.}},
\bauthor{\bsnm{Chen}, \binits{S.A.}},
\bauthor{\bsnm{Ye}, \binits{P.}}:
\batitle{{Entanglement, non-{H}ermiticity, and duality}}.
\bjtitle{SciPost Phys.}
\bvolume{11},
\bfpage{003}
(\byear{2021}).
\doiurl{10.21468/SciPostPhys.11.1.003}
\end{barticle}
\endbibitem

\bibitem{hatsugai2006}
\begin{barticle}
\bauthor{\bsnm{Hatsugai}, \binits{Y.}}:
\batitle{Quantized {B}erry phases as a local order parameter of a quantum
  liquid}.
\bjtitle{J. Phys. Soc. Jpn.}
\bvolume{75}(\bissue{12}),
\bfpage{123601}
(\byear{2006}).
\doiurl{10.1143/JPSJ.75.123601}
\end{barticle}
\endbibitem

\bibitem{liang2013}
\begin{barticle}
\bauthor{\bsnm{Liang}, \binits{S.-D.}},
\bauthor{\bsnm{Huang}, \binits{G.-Y.}}:
\batitle{Topological invariance and global {B}erry phase in non-{H}ermitian
  systems}.
\bjtitle{Phys. Rev. A}
\bvolume{87}(\bissue{1}),
\bfpage{012118}
(\byear{2013}).
\doiurl{10.1103/PhysRevA.87.012118}
\end{barticle}
\endbibitem

\bibitem{zak1989}
\begin{barticle}
\bauthor{\bsnm{Zak}, \binits{J.}}:
\batitle{Berry's phase for energy bands in solids}.
\bjtitle{Phys. Rev. Lett.}
\bvolume{62},
\bfpage{2747}--\blpage{2750}
(\byear{1989}).
\doiurl{10.1103/PhysRevLett.62.2747}
\end{barticle}
\endbibitem

\bibitem{cardano2017}
\begin{barticle}
\bauthor{\bsnm{Cardano}, \binits{F.}},
\bauthor{\bsnm{D'Errico}, \binits{A.}},
\bauthor{\bsnm{Dauphin}, \binits{A.}},
\bauthor{\bsnm{Maffei}, \binits{M.}},
\bauthor{\bsnm{Piccirillo}, \binits{B.}},
\bauthor{\bparticle{de} \bsnm{Lisio}, \binits{C.}},
\bauthor{\bsnm{De~Filippis}, \binits{G.}},
\bauthor{\bsnm{Cataudella}, \binits{V.}},
\bauthor{\bsnm{Santamato}, \binits{E.}},
\bauthor{\bsnm{Marrucci}, \binits{L.}},
\bauthor{\bsnm{Lewenstein}, \binits{M.}},
\bauthor{\bsnm{Massignan}, \binits{P.}}:
\batitle{Detection of {Z}ak phases and topological invariants in a chiral
  quantum walk of twisted photons}.
\bjtitle{Nat. Commun.}
\bvolume{8}(\bissue{1}),
\bfpage{15516}
(\byear{2017}).
\doiurl{10.1038/ncomms15516}
\end{barticle}
\endbibitem

\bibitem{shen2018}
\begin{barticle}
\bauthor{\bsnm{Shen}, \binits{H.}},
\bauthor{\bsnm{Zhen}, \binits{B.}},
\bauthor{\bsnm{Fu}, \binits{L.}}:
\batitle{Topological band theory for non-{H}ermitian {H}amiltonians}.
\bjtitle{Phys. Rev. Lett.}
\bvolume{120}(\bissue{14}),
\bfpage{146402}
(\byear{2018}).
\doiurl{10.1103/PhysRevLett.120.146402}
\end{barticle}
\endbibitem

\bibitem{zhang2019}
\begin{barticle}
\bauthor{\bsnm{Zhang}, \binits{X.Z.}},
\bauthor{\bsnm{Song}, \binits{Z.}}:
\batitle{Partial topological {Z}ak phase and dynamical confinement in a
  non-{H}ermitian bipartite system}.
\bjtitle{Phys. Rev. A}
\bvolume{99},
\bfpage{012113}
(\byear{2019}).
\doiurl{10.1103/PhysRevA.99.012113}
\end{barticle}
\endbibitem

\bibitem{brody2013}
\begin{barticle}
\bauthor{\bsnm{Brody}, \binits{D.C.}}:
\batitle{Biorthogonal quantum mechanics}.
\bjtitle{J. Phys. A: Math. Theor.}
\bvolume{47}(\bissue{3}),
\bfpage{035305}
(\byear{2013}).
\doiurl{10.1088/1751-8113/47/3/035305}
\end{barticle}
\endbibitem

\bibitem{song2012a}
\begin{barticle}
\bauthor{\bsnm{Song}, \binits{H.F.}},
\bauthor{\bsnm{Rachel}, \binits{S.}},
\bauthor{\bsnm{Flindt}, \binits{C.}},
\bauthor{\bsnm{Klich}, \binits{I.}},
\bauthor{\bsnm{Laflorencie}, \binits{N.}},
\bauthor{\bsnm{Le~Hur}, \binits{K.}}:
\batitle{Bipartite fluctuations as a probe of many-body entanglement}.
\bjtitle{Phys. Rev. B}
\bvolume{85}(\bissue{3}),
\bfpage{035409}
(\byear{2012}).
\doiurl{10.1103/PhysRevB.85.035409}
\end{barticle}
\endbibitem

\bibitem{acharya2022}
\begin{barticle}
\bauthor{\bsnm{Acharya}, \binits{A.P.}},
\bauthor{\bsnm{Chakrabarty}, \binits{A.}},
\bauthor{\bsnm{Sahu}, \binits{D.K.}},
\bauthor{\bsnm{Datta}, \binits{S.}}:
\batitle{Localization, $\mathcal{PT}$ symmetry breaking, and topological
  transitions in non-{H}ermitian quasicrystals}.
\bjtitle{Phys. Rev. B}
\bvolume{105}(\bissue{1}),
\bfpage{014202}
(\byear{2022}).
\doiurl{10.1103/PhysRevB.105.014202}
\end{barticle}
\endbibitem

\bibitem{ashida2017}
\begin{barticle}
\bauthor{\bsnm{Ashida}, \binits{Y.}},
\bauthor{\bsnm{Furukawa}, \binits{S.}},
\bauthor{\bsnm{Ueda}, \binits{M.}}:
\batitle{Parity-time-symmetric quantum critical phenomena}.
\bjtitle{Nat. Commun.}
\bvolume{8},
\bfpage{15791}
(\byear{2017}).
\doiurl{10.1038/ncomms15791}
\end{barticle}
\endbibitem

\end{thebibliography}



\end{document}